\documentclass[aps,superscriptaddress,preprint,prl]{revtex4-1}
\usepackage{marvosym}
\usepackage{graphicx, subfigure}         
\usepackage{amsmath}                     
\usepackage{bm}                     
\usepackage{color}
\usepackage{wrapfig}
\usepackage{mathtools}
\usepackage{tikz}  
\usetikzlibrary{arrows,shapes,trees,positioning}  
\usepackage{soul}

\usepackage{comment} 

\begin{document}


\title{Analysis of the linear relationship between asymmetry and magnetic moment at the M-edge of $3d$ transition metals}

\author{Somnath Jana}
\affiliation{Dept.\ of Physics and Astronomy, Uppsala University, Box 516, SE-75120 Uppsala, Sweden}
\author{R. S. Malik}
\affiliation{Dept.\ of Physics and Astronomy, Uppsala University, Box 516, SE-75120 Uppsala, Sweden}
\author{Yaroslav O. Kvashnin}
\affiliation{Dept.\ of Physics and Astronomy, Uppsala University, Box 516, SE-75120 Uppsala, Sweden}
\author{Inka L. M. Locht}
\affiliation{Dept.\ of Physics and Astronomy, Uppsala University, Box 516, SE-75120 Uppsala, Sweden}
\author{R. Knut}
\affiliation{Dept.\ of Physics and Astronomy, Uppsala University, Box 516, SE-75120 Uppsala, Sweden}
\author{R. Stefanuik}
\affiliation{Dept.\ of Physics and Astronomy, Uppsala University, Box 516, SE-75120 Uppsala, Sweden}
\author{Igor Di Marco}
\affiliation{Dept.\ of Physics and Astronomy, Uppsala University, Box 516, SE-75120 Uppsala, Sweden}
\affiliation{Asia Pacific Center for Theoretical Physics, Pohang, Gyeonbuk 790-784, Republic of Korea}
\affiliation{Department of Physics, POSTECH, Pohang, Gyeonbuk 790-784, Republic of Korea}
\author{A.N. Yaresko}
\affiliation{Max-Planck-Institut f{\"u}r Festk{\"o}rrperforschung,
Heisenbergstrasse 1, D-70569 Stuttgart, Germany}
\author{Martina Ahlberg}
\affiliation{Dept.\ of Physics, University of Gothenburg, Gothenburg, Sweden}
\author{Raghuveer Chimata}
\affiliation{Dept.\ of Physics and Astronomy, Uppsala University, Box 516, SE-75120 Uppsala, Sweden}
\affiliation{Argonne National Lab. Lemont IL 60439, USA}
\author{Marco Battiato}
\affiliation{Nanyang Technological University, 21 Nanyang Link, Singapore, Singapore}
\author{Johan S\"oderstr\"om}
\affiliation{Dept.\ of Physics and Astronomy, Uppsala University, Box 516, SE-75120 Uppsala, Sweden}
\author{Olle Eriksson}
\affiliation{Dept.\ of Physics and Astronomy, Uppsala University, Box 516, SE-75120 Uppsala, Sweden}
\affiliation{School of Science and Technology, \"Orebro University, \"Orebro, Sweden}
\author{Olof Karis}
\affiliation{Dept.\ of Physics and Astronomy, Uppsala University, Box 516, SE-75120 Uppsala, Sweden}


\begin{abstract}

The magneto-optical response of Fe and Ni during ultrafast demagnetization is studied experimentally and theoretically. We have performed pump-probe experiments in the transverse magneto-optical Kerr effect (T-MOKE) geometry using photon energies that cover the M-absorption edges of Fe and Ni between 40 to 72 eV. The asymmetry was detected by measuring the reflection of light for two different orientations of the sample magnetization.
Density functional theory (DFT) was used to calculate the magneto-optical response of different magnetic configurations, representing different types of excitations: long-wavelength magnons, short wavelength magnons, and Stoner excitations.
In the case of Fe, we find that the calculated asymmetry is strongly dependent on the specific type of magnetic excitation. Our modelling also reveals that during remagnetization Fe is, to a reasonable approximation, described by magnons, even though small non-linear contributions could indicate some degree of Stoner excitations as well.
In contrast, we find that the calculated asymmetry in Ni is rather insensitive to the type of magnetic excitations.
However, there is a weak non-linearity in the relation between asymmetry and the off-diagonal component of the dielectric tensor, which does not originate from the modifications of the electronic structure.
Our experimental and theoretical results thus emphasize the need of considering a coupling between asymmetry and magnetization that may be more complex that a simple linear relationship. This insight is crucial for the  microscopic interpretation of ultrafast magnetization experiments.

\end{abstract} 

\maketitle

\section{Introduction}
Since the first experimental observation of ultrafast demagnetization in Ni~\cite{PhysRevLett.76.4250}, ultrafast magnetodynamics has attracted significant interest, due to the perspective of controlling the magnetization on sub-picosecond timescales~\cite{Koopmans05,Stamm07,Krauss09,Bigot09,Koopmans10,Carva11,Rudolf12,graves13,Carva13,Battiato14,Schellekensnatcomm14,Tows15,Elyasi16,Nenno16}. These studies have led to the discovery of novel effects such as all-optical-switching~\cite{Stanciu07,RevModPhys.82.2731} and ultrafast spin currents~\cite{Malinowski08,Battiato10,Melnikov11,Eschenlohrnatmater2013,Battiato16}. 
Probing the magnetic state in condensed matter by using light has been a widespread approach for many decades. This can be achieved in either reflectivity measurements, based on the magneto-optical Kerr effect (MOKE), or in transmission measurements using the Faraday effect or magnetic circular dichroism. 
The relation between the magneto-optical response of the material in the sub-picosecond timescale and its instantaneous magnetization has been investigated in the past for various geometries~\cite{PhysRevB.12.5016,Koopmans00,Guidoni02,Bigot04,Carpene08,Zhang09,Carpene2015FeNotStoner,PhysRevX.2.011005,PhysRevB.92.064403}, but, as illustrated below, no clear conclusion was reached and the topic still remains controversial. 
Using light as both pump and probe provides the possibility of studying ultrafast magnetic processes in the femtosecond time regime. Typically, an intense ultrashort (tens of femtoseconds) laser pulse is used to excite the magnetic sample and then the response is measured by another temporally short but less intense optical pulse, so that the latter has an insignificant effect on the magnetic state. Most ultrafast magnetization studies have been performed by using light in the visible regime due to it being easily accessible.
However, during the last few decades, high photon energies have become available through high harmonic generation (HHG) sources reaching extreme ultraviolet (EUV) and by synchrotrons facilities providing energies in the soft x-ray regime.
The main advantage of using EUV or higher photon energies is that one can reach absorption edges that provide higher magnetic contrast and elemental selectivity. 
With the access to  HHG sources, element-specific magnetization dynamics becomes possible in lab-based setups, where the magnetic state is probed in the transverse MOKE (T-MOKE) geometry using linearly polarized XUV. 
The T-MOKE geometry is highly advantageous for high energy photons as circularly polarized light is not easily accessible for high harmonics and polarimetry is not easily performed due to the limited availability of optical components at high photon energies. 
The T-MOKE signal is commonly assumed to be approximately proportional to the magnetization of the probed sample (see e.g.\ \cite{Mathias2012,Mathias2013164}). Since the signal is strongly enhanced near the absorption edge of a given magnetic element, the proportionality is expected to be dominated by the local magnetization of that element at the resonant energies \cite{Mathias2013164,Mathias2012,PhysRevX.2.011005,Rudolf12}.

Similar to previous studies \cite{Mathias2013164,Mathias2012,PhysRevX.2.011005,Rudolf12}, we have measured the reflected intensity of p-polarized light at angles close to the Brewster angle. The measurements were made for two different magnetization directions (yielding $I_+$ and $I_-$) and the asymmetry, $A(E)$ in the measurement was defined as \cite{handbookMagMat, PhysRevX.2.011005,Mathias2013164}: 

\begin{equation}
  A(E)= \frac{I_+(E)-I_-(E)} {I_+(E)+I_-(E)}. 
 \label{prop-rel}
\end{equation}
Assuming that the off-diagonal components to the dielectric tensor is small compared to the Fresnel coefficient, one may conclude that
\begin{equation}
A(E) \approx -\tan{(2\theta)}
\left( \frac{\delta(E) \epsilon^1_{xy}(E) - \beta(E) \epsilon^2_{xy}(E)}{\delta^2(E) + \beta^2(E)} \right),
\label{prop-rel2}
\end{equation}
where $\epsilon^1_{xy}$ and $\epsilon^2_{xy}$ are the real and imaginary parts of the off-diagonal components of the dielectric tensor $\bm{\epsilon}$, respectively, and $\theta$ is the angle of incidence relative to the surface normal. In this equation $\delta(E)$ and $\beta(E)$ are components of the refractive index that depend on the energy of the light. Owing to the Onsager relationship, $\epsilon_{xy}$ changes sign when the magnetic moment of the sample is reversed. It thus follows from Eq.\ \ref{prop-rel2} that the asymmetry, $A(E)$ is an odd function in the magnetization {\bf M}, for any energy $E$. Often a linear relationship is assumed, but one should note that this is not always a valid approximation, as will be demonstrated here. 
The reason can be twofold, either there are higher order contributions in {\bf M} influencing $\epsilon_{xy}$ (via changes in the electronic structure in the excited states) or additional contributions to $A(E)$ which are not accounted for by the approximated Eq.~\ref{prop-rel2}.
The way this is analysed here is done by making an assumption that $\epsilon_{xy} \propto {\bf M}$. From Eq.\ \ref{prop-rel2} it then follows that 
\begin{eqnarray}
A(E) \approx K(E) {\bf M},
\label{prop-rel3}
\end{eqnarray}
where $K(E)$ is an energy-dependent proportionality constant that involves the different terms of Eq.~\ref{prop-rel2} (e.g.\ $\delta$ and $\beta$). 
If the relationship of Eq.~\ref{prop-rel3} holds, one should be able to follow the asymmetry of the reflected light for two different photon energies, E and E$^{\prime}$ and the ratios in asymmetry would be $\frac{K(E)}{K(E^{\prime})}$, i.e. a constant that does not depend on {\bf M}. Any deviation from this behaviour in e.g.\ a pump-probe experiment, must indicate that higher order contributions in {\bf M} influence $\epsilon_{xy}$, demonstrating complexities in the band structure. In this study, we will undertake such an analysis and demonstrate that one can use this relationship between $A(E)$ and {\bf M} to draw conclusions about the microscopic mechanisms of the magnetization dynamics of such experiments. 
It is important to stress that non-magnetic contributions due to the transient variation of the refractive index by the non-equilibrium hot-electron distribution are negligible for the chosen experimental geometry~\cite{PhysRevX.2.011005}.


Ultrafast demagnetization has previously been studied in a Co film, by means of measuring the asymmetry from the T-MOKE signal at M edge \cite{Turgut2016StonerDemagnetization,Zusin2018DirectScanning}. The results conclude that both ultrafast magnon generation and transient reduction of the exchange splitting are responsible for the ultrafast demagnetization. Analysis at two different pump-probe delays suggests that the ultrafast magnon generation is dominant during subpicosecond time scales, while the transient reduction of the exchange splitting persists for several picoseconds. This is rather counter-intuitive from the total energy consideration. This scenario was supported by an independent photoemission study~\cite{Eiche1602094}.
The pump-probe response of Co was also investigated by atomistic spin-dynamics simulations \cite{Chimata2017_Co_CoMn}, in which the relevant information is obtained from \textit{ab initio} electronic structure theory (see also \cite{UPP_ASD_Book}), which argued that the magnetic response should be dominated by transverse spin-disorder, following the two-temperature model\cite{Koopmans10}.
 
In a separate work on Fe, by means of magneto-optical and reflectivity spectroscopy, transversal spin fluctuation was found to be the dominant mechanism rather than a change in exchange splitting \cite{Carpene2015FeNotStoner}, which is consistent with results from atomistic spin-dynamics simulations, although the lower range for appropriate time scales for such simulations is of order of a few hundreds fs \cite{Chimata2012remag}.
Recently, the problem of laser-induced demagnetization has been addressed by time-dependent density functional theory (TD-DFT)\cite{PhysRevLett.52.997}. This is a rigorous approach, which in principle describes the evolution of the electron density in the first moments after the optical irradiation.
Practical implementations are naturally associated with approximations, and the results of TD-DFT calculations have to be critically examined and compared to experiments. 
The results obtained for three elemental ferromagnets  (Fe, Co, Ni) reveal similar trends of dynamics, where the demagnetization is primarily caused by the spin-flip electron excitation \cite{Krieger15,PhysRevLett.119.107203}. Short wavelength magnons are predicted to be irrelevant at the ultrashort timescales, but the predictive power of these studies is limited by the small size of computationally treatable supercells. In addition, it is reported that the demagnetization is enhanced at the surface of a material as compared to the bulk \cite{Krieger_2017}.

The examples above serve to illustrate that a firm theoretical understanding of pump-probe experiments in the fs time-scale, is still evolving and that additional analysis and measurements are needed. To this end it is important to clarify the proportionality relations described in Eq.~\ref{prop-rel}, something we attempt to do here. In this work, we have studied the magnetic asymmetry of the ferromagnetic films (Fe, Ni) in T-MOKE geometry at probing photon energy ranging from 40 - 72 eV. 
The dynamics of the observed magnetic signal at different photon energies is compared to \textit{ab initio} derived theoretical spectra, corresponding to several possible types of magnetic excitations. 
It is found that Fe exhibits strong non-linear effects in the energy dependent asymmetry during the demagnetization. However, during the remagnetization mainly a linear response is found, which indicates that the system contains mainly longer wavelength magnons. 
Ni exhibits a more or less proportional relation between asymmetry and magnetization for all types of excitations considered in our modelling. This is also reflected in the experimental data, that only show a weak energy dependence of the asymmetry. We note that there is an additional non-linearity between A and $\epsilon_{xy}$, which is more pronounced in Ni than in Fe.

\begin{figure*}[]
     \centering
     \includegraphics[width=\textwidth]{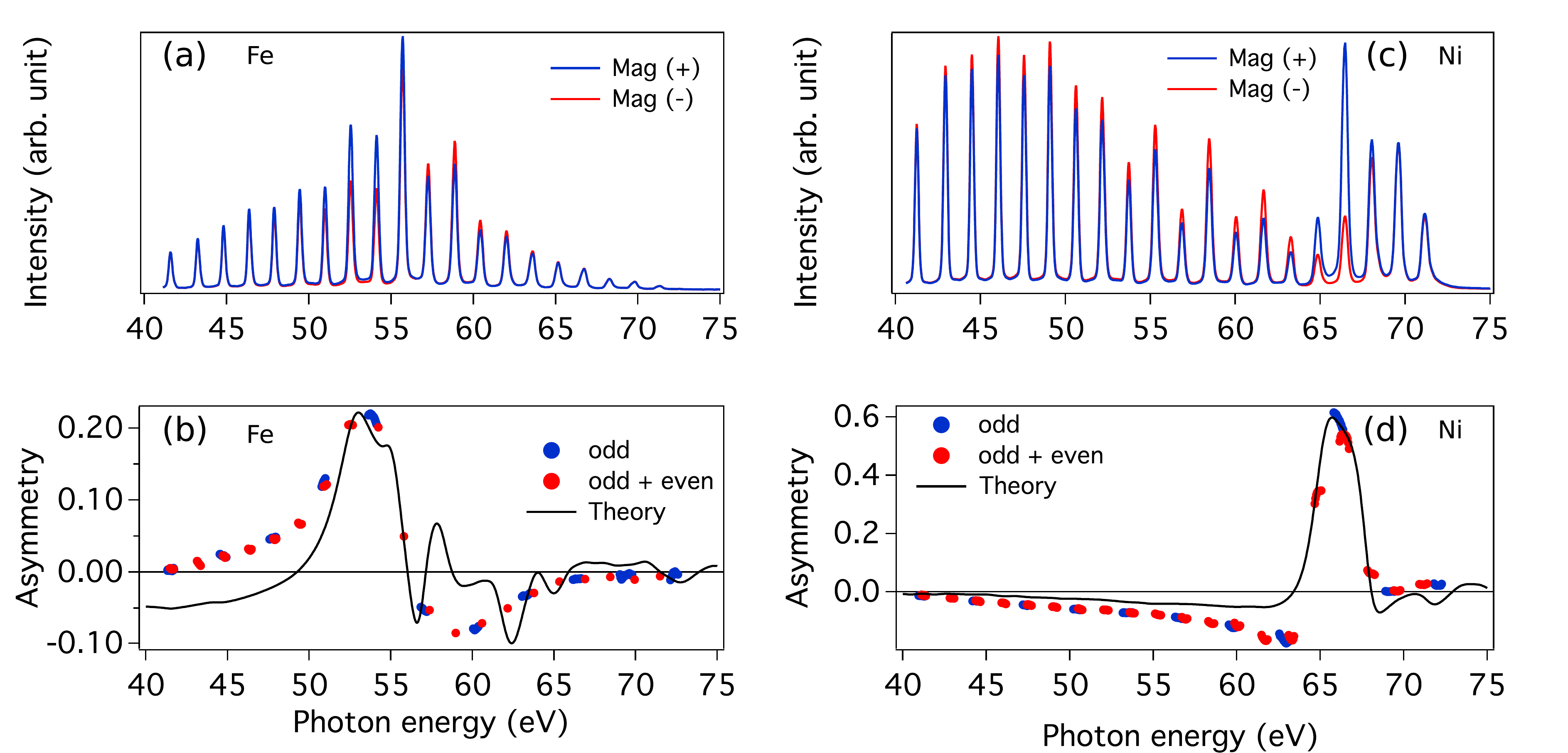} 
\caption{(a) The reflected harmonic intensities $I_{\pm}(E)$ for Fe. (b) The corresponding magnetic asymmetry for Fe, using "odd" and "odd and even" harmonics as explained in the text. (c) The reflected harmonic intensities $I_{\pm}(E)$ for Ni.  (b) The corresponding magnetic asymmetry for Ni, using "odd" and "odd and even" harmonics as explained in the text. Ground state theoretical asymmetry spectra are plotted in (b) and (d). See text for computational details.}
\label{fig:Asymexp}
\end{figure*}

\section{Experiment}\label{sec:experiment}
\subsection{Experimental setup}\label{sec:Experimentalsetup}
The experiments are performed at the HELIOS lab, Uppsala University, Sweden. The near-infrared (NIR) pump pulse has a wavelength of $\sim 800$~nm ($\sim 1.5$~eV) and a pulse length of $\sim 35$~fs. The broadband EUV probe pulse has energies between $40 - 72$~eV and a pulse length of $\sim 20$~fs \cite{Plogmaker2015,STEFANUIK201833,Jana2017}. 
Since the NIR pump and the EUV pulse are generated from the same  laser pulse, the temporal jitter between the NIR and EUV is eliminated and hence the temporal resolution is determined by the pulse widths. 
The measurement geometry is described in detail in Ref.\ \cite{Jana2017}. Both the $p$-polarized EUV and the NIR are focused on the sample at about 45$^{\circ}$ incidence angle, i.e.\ near the Brewster angle for EUV photons, which provides a strong magneto-optical signal. The reflected EUV probe is dispersed by a subsequent spectrometer, while the reflected NIR pump is absorbed by a 200 $\mu$m thick Al foil~\cite{Jana2017}. In the T-MOKE geometry, the sample is magnetized perpendicularly to the plane of incidence of the incoming $p$-polarized light. The reflectivity for opposite magnetization directions ($\pm {\bf M}$) is measured as a function of the harmonic energy E, and is labeled $I_{\pm}(E)$. A measured energy dependent background (dark noise in the detector system) has been subtracted from all spectra. For time-resolved pump-probe measurements, the spectra are comprised of odd harmonics of the fundamental light, thus having a splitting of $\sim 3.1\,\mbox{eV}$. For the static case, HHG spectra containing all  harmonics (even and odd), were obtained by mixing the fundamental light with its second harmonic in the HHG process \cite{PhysRevA.82.033410}, as shown in Fig.~\ref{fig:Asymexp}(a and c).
  
From a practical point of view, theory of the asymmetry function has to use a slightly modified version of Eq.~1, i.e.\  
\begin{equation}\label{eq:asymdiff}
A(E) = \frac{I_+(E) - I_- (E)}{I_+(E) + I_- (E)+\Gamma} \:.
\end{equation}
where Eq.\ 1 has been augmented by a constant $\Gamma$ in the denominator, to account for effects that are not included in the theoretical calculations, e.g.\ the background signal. We will discuss this point in more detail in the Theory section. 

Two samples with elemental films of Fe and Ni, respectively, were used in this investigation. A 100 nm Ni film was deposited on a Si wafer by means of magnetron sputtering. Upon exposure to air this sample will form a native oxide layer of  about 1 nm thickness \cite{holloway1974_NiO,tyuliev1999_NiO}. Similarly, a 100 nm Fe film was sputtered on a Si wafer and subsequently capped with a 5 nm Cu layer to prevent oxidation of the Fe film. 

\subsection{Experimental results}\label{sec:Experimental results}
\begin{figure}
     \centering
     \includegraphics[width=1.0\columnwidth]{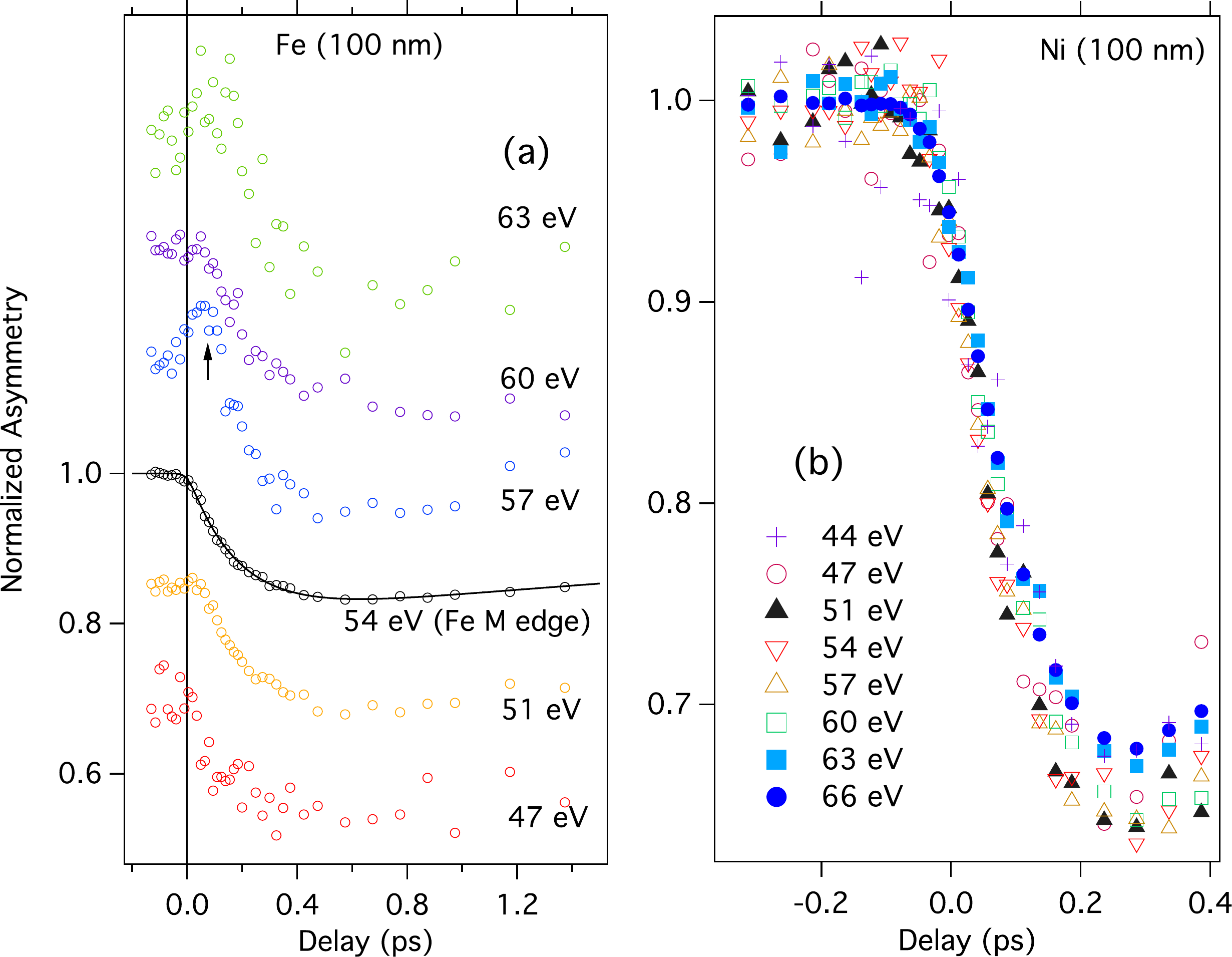} 
\caption{(a) Measured asymmetry for the bcc Fe sample at different harmonic energies, as a function of delay time between pump and probe. There are  strong irregularities in the energy dependent asymmetry data for Fe,  highlighted by the arrow (63, 60, 57, and 51 eV). All curves except at 54 eV have been shifted vertically for clarity. For the 54 eV harmonic, we also show a representative fit to the data (solid black line).
(b) Measured asymmetry for the fcc Ni sample at different harmonic energies, as a function of delay time between pump and probe. No irregularities, similar to those found for Fe, are found for Ni.}
\label{fig:Demagcurvexp}
\end{figure}

The difference in arrival time between the pump and probe light is controlled by a delay stage.
If the probe pulse arrives before the pump pulse, the sample will be fully magnetized, resulting in a large asymmetry. The measured asymmetry $A(E)$, for fully magnetized Fe and Ni is reported in Figs.~\ref{fig:Asymexp}(b) and \ref{fig:Asymexp}(d), respectively. As can be observed in Fig.~\ref{fig:Asymexp}(b), the asymmetry of Fe is distributed over the whole available spectrum of harmonics, but most strongly peaked at one harmonic energy. The same effect can be observed for the Ni sample, in Fig.~\ref{fig:Asymexp}(d), but the main peak is now shifted with about $\sim$ 12~eV to higher photon energy. Since the energy splitting between odd harmonics is relatively large ($\sim$ 3.1 eV), we also present the asymmetry obtained with spectra containing both even and odd harmonics, giving a splitting of $\sim$ 1.55 eV. This provides a more detailed experimental determination of the asymmetry and shows that the odd harmonics, which are used for our time-resolved studies, capture the peak asymmetry of both Fe and Ni at $\sim$ 54 eV and $\sim$ 66 eV, respectively. 
If the probe arrives after the pump, the sample demonstrates a decreased asymmetry due to the ultrafast demagnetization process. This is illustrated in Figs.~\ref{fig:Demagcurvexp}(a) and (b), where the asymmetries of Fe and Ni are reported as a function of the delay between pump and probe. At $t=0$~ps, the pump and probe pulses are incident on the sample simultaneously. Within a time delay of less than 1~ps the sample shows a strong reduction of the asymmetry, i.e. a strong demagnetization. After the initial reduction, the asymmetry starts returning to its initial value, which indicates a recovery of the sample magnetization on a longer timescale (10s of ps).

While it is certain that the asymmetry up to a large degree, reflects the sample magnetization, their precise relationship is rather complex. If the asymmetry were proportional to the magnetization, we would expect the asymmetry to be independent of the photon energy of the probe pulse during the demagnetization and remagnetization process. The magnetic asymmetry of Fe during the demagnetization process is shown in Fig.~\ref{fig:Demagcurvexp}(a) for six different photon energies. For the curve corresponding to the harmonic with the peak asymmetry ($\sim$54~eV), the reduction of the asymmetry starts earlier than for most of the curves at both higher and lower photon energies. This illustrates the complexity of identifying a relationship between magnetic asymmetry and magnetization. Analogous data for Ni is reported in Fig.~\ref{fig:Demagcurvexp}(b), but, unlike for Fe, the curves show much less variation with the photon energy.

\begin{figure}
    \centering
    \includegraphics[width=0.70\columnwidth]{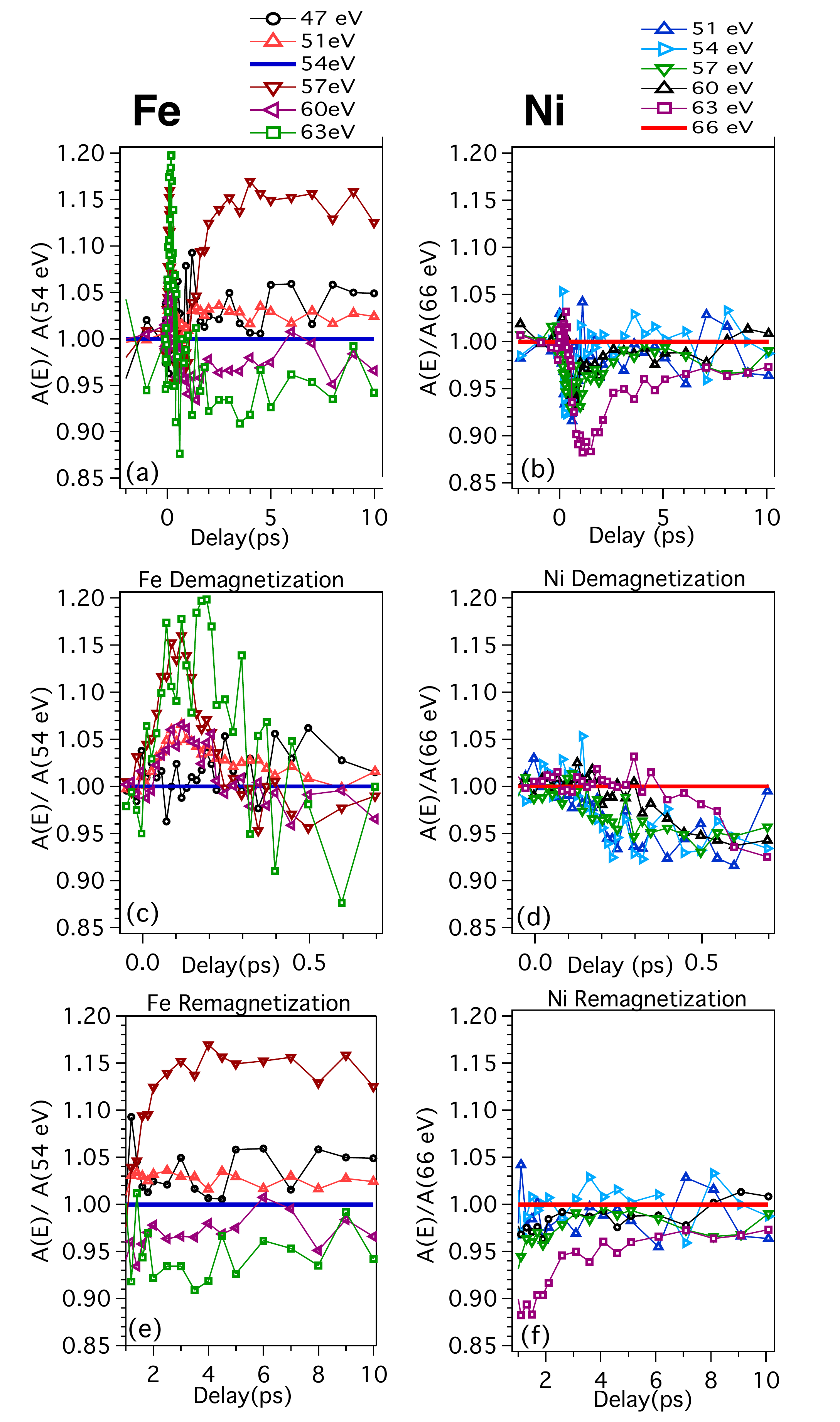}
    \caption{The measured magnetisation-asymmetry test ratio (MAT ratio - for details see text) of bcc Fe and fcc Ni as a function of time delay 
    is shown in (a-b). The measured data are shown in two time-domains that represent the demagnetization (c-d) as well as the remagnetization (e-f) process.}
    \label{fig:MDT_ratio_fig}
\end{figure}

According to the discussion around Eqns.\ \ref{prop-rel}-\ref{prop-rel3}, one would expect that if the measured asymmetry is proportional to the magnetization (Eq.\ \ref{prop-rel3}),  the ratio of asymmetries detected at two different energies would be independent on the magnetic moment. In a pump-probe experiment, this be manifested as a time-independent ratio. In order to investigate this, we have taken the ratio between all measured curves in Fig.\ \ref{fig:Asymexp}, with respect to the measured asymmetry at energies close to the transition edge (54 eV for bcc Fe and 66 eV for fcc Ni). Since the asymmetry for each harmonic is independently normalized with respect to its $t=0$ value, the ratios lie close to unity. We have plotted these ratios in Fig.~\ref{fig:MDT_ratio_fig} and we will refer to such ratios as the magnetisation-asymmetry test ratio (MAT ratio), both for the demagnetization (0-0.75 ps) and the remagnetization processes  $(t\geq 0.75 \mbox{ps})$. It may be noted from the figure that for fcc Ni the MAT ratio is much more constant than for bcc Fe, suggesting that for fcc Ni the measured asymmetry indeed does reflect the magnetic state to a high accuracy. However, for bcc Fe the situation is much more complex. For this element, the demagnetization process shows a MAT ratio that clearly casts doubt on the asymmetry being proportional to the magnetic moment, since MAT ratios taken at many different energies are far from being time-independent. The remagnetization process of bcc Fe at pump-probe delays longer than 2 ps is more well behaved, as suggested by fact that the corresponding MAT curves are close to constant for $t>2\, \mbox{ps}$.

\section{Theory}\label{sec:theory}
\subsection{Computational details}\label{sec:theorexpcorr}

 To analyze the results in Figs.\ \ref{fig:Demagcurvexp} and \ref{fig:MDT_ratio_fig}, we have performed ab-initio calculations of the relationship between the magnetic asymmetry and the magnetization for Fe and Ni. As recently demonstrated~\cite{PhysRevB.92.064403}, calculations do not need to include transient effects explicitly, but can be performed in the quasi-static limit, where the sample magnetization can be treated as a constraint.
 In our approach, the magnetic asymmetry is calculated in two steps. 
First, the dielectric tensor is calculated for the bulk material from first principles by means of DFT.
Thereafter the Fresnel equations, which define the optical response of the material are solved numerically~\cite{PhysRevLett.104.187401,YEH198041} for a chosen sample thickness. The resulting reflectivities for the two magnetization directions are then used to obtain the magnetic asymmetry via Eq.~\ref{eq:asymdiff}, 
in which $\Gamma$ is a constant-energy background to take into account extrinsic contributions to the reflectivity, such as those arising from impurities, substrate or surface oxide layer.
We have chosen $\Gamma$ equal to 0.0052 in case of Fe and 0.004 for  Ni respectively, in order to match the peak asymmetry between theoretical and experimental results. The same values of $\Gamma$ were then used to simulate the asymmetry for the partially demagnetized configurations.

\begin{figure}
     \centering
     \includegraphics[width=1.0\columnwidth]{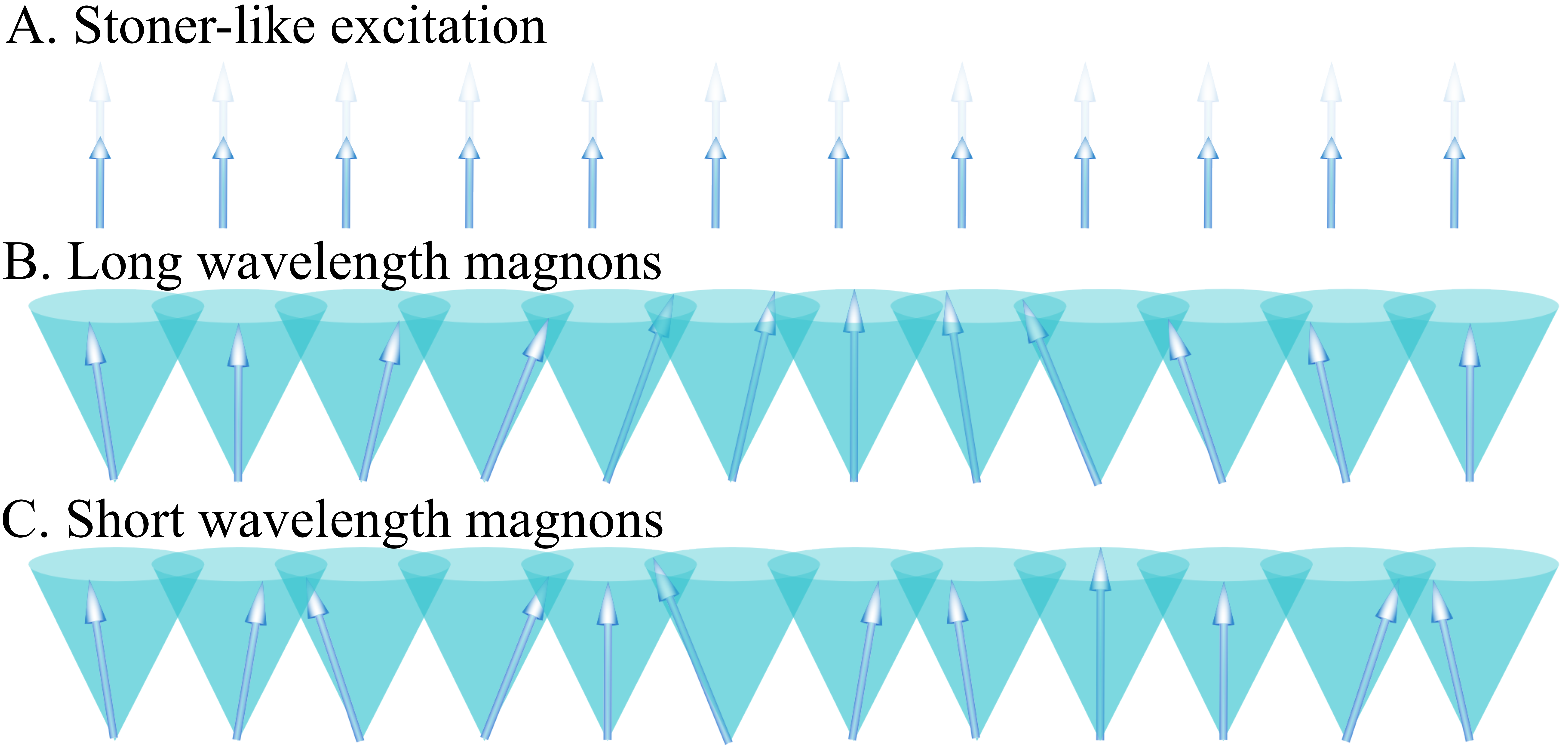} 
\caption{Illustration of different microscopic states with decreased total magnetization.}
\label{fig:demagcases}
\end{figure}

The DFT calculations were done using the full-potential linearized augmented plane wave (FP-LAPW) method as implemented in the ELK code~\cite{elk_website}. 
Both materials were considered in their ground state structures with experimental lattice parameters equal to 5.42 a.u. and 6.66 a.u., for Fe and Ni, respectively. Exchange and correlation effects were treated with the PBE functional \cite{pbe}. 
The radii of the muffin-tin spheres $R_{MT}$ were set to 2.0 a.u.\ for both Fe and Ni and $3p$ states were treated as a part of the valence band. 
The k-point grid of 30$^3$ k-points was employed together with a fairly large amount of empty states (18) in order to ensure the convergence of the optical properties.
The dielectric tensor was calculated from the obtained band structure using standard Kubo-Greenwood formalism. The dielectric tensor was subsequently convoluted by a Gaussian of 1.2 eV to mimic the experimental asymmetries. The corresponding calculated ground state asymmetries are presented in Fig.~\ref{fig:Asymexp}(b) and (d) for Fe and Ni, respectively. 
An energy shift of 2.0 eV for Fe and 2.2 eV for Ni, respectively, was applied to the theoretical asymmetry spectra in order to align them to the experimental data. The origin of these shifts has been very recently identified in a mix of many-body effects, which are only roughly described in Kohn-Sham DFT, and local field effects~\cite{PhysRevLett.122.217202}.
As is clear from Fig.~\ref{fig:Asymexp}(b) and (d), the calculations and experiments are found to agree rather well. Theory and experiment show a smooth behavior below the peak asymmetry for both Fe and Ni. Also, large asymmetries are found up to 10 eV above the peak asymmetry for Fe, even though theory indicates sharper structures in the asymmetry that are not observed experimentally. For Ni, both theory and experiment show that the asymmetry is strongly diminished at higher energies.
\begin{figure}
  \includegraphics[width=0.9\columnwidth]{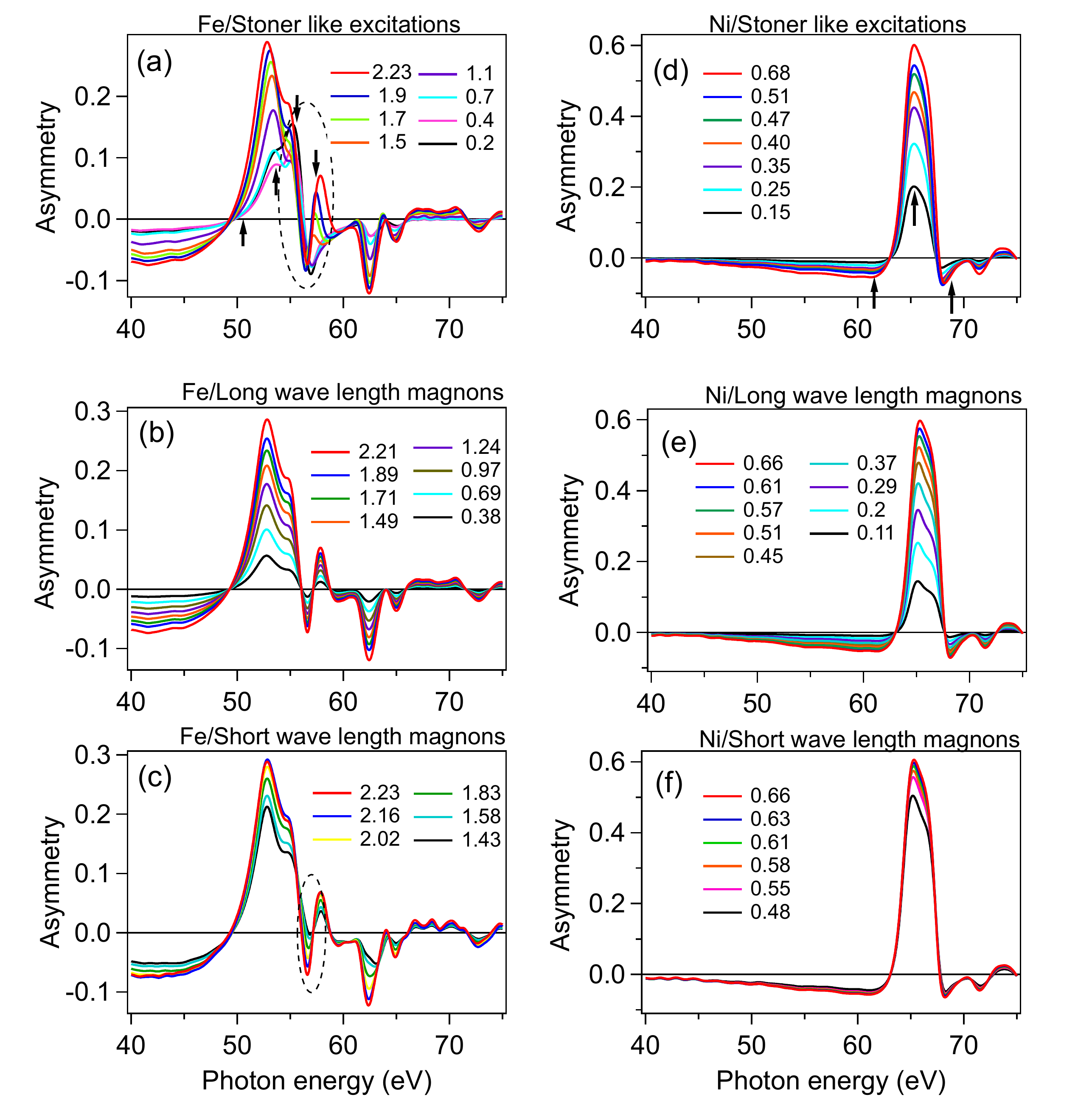}
\caption{Theoretical asymmetries for three different types of magnetic excitations, namely, Stoner-like excitation where moments are decreased collinearly, long wavelength magnon excitations (gradually tilted magnetic moments) and a high density of short wavelength magnons (randomly tilted magnetic moments) are plotted for different magnetization in panels (a), (b), (c) for Fe and (d), (e), (f) for Ni respectively. The value of the reduced magnetic moment for each configuration is given in the upper right part of each plot in units of $\mu_B$. The region indicated by a dashed ellipse is discussed in the text. The arrows indicate the energies used to calculate the real and imaginary parts of $\epsilon_{xy}$, presented in Fig.\ \ref{fig:theo_resonance_eps}}.
\label{fig:theo_asym_spectra}
\end{figure}
As mentioned above, to investigate the relation between the magnetic asymmetry and the magnetization of the sample, we have treated the system in the quasi-static limit~\cite{PhysRevB.92.064403}. In this regime, the local magnetic moments are well-defined and can be used to classify all possible microstates of the system. For our purposes we have to consider only those microstates that are compatible with the decreased value of the $z$-projection of the magnetization. These states can be grouped in three families, as illustrated in Fig.~\ref{fig:demagcases}. In panel A the total magnetization is decreased by shrinking the length of all atomic magnetic moments (corresponding to Stoner excitations), in panel B the total magnetization is decreased by exciting long wavelength spin waves, and in panel C the atomic magnetic moments are tilted from the $\hat{z}$-axis over an angle $\Theta$ in random directions, thus representing a high density of short wavelength magnons.
For these possible scenarios, we have calculated the dielectric tensor and the corresponding asymmetry for a prescribed value of {\bf M}. 
Each microscopic picture was modeled in the following way:

\textit{Stoner like excitations:}
Collinearly decreased atomic magnetic moments. The ferromagnetic atomic magnetic moments, projected onto muffin-tin sphere, were shrunk to a certain value using constrained DFT.

\textit{Long wavelength magnon excitations:}
This can also be though of as a gradual tilting of atomic magnetic moments. The dielectric tensor calculated for the equilibrium ferromagnetic state was tilted by an angle $\Theta$ from the $\hat{z}$-axis, rotated by an angle $\phi$ around this axis and then averaged over all angles $0 \le \phi \le 2\pi$. This implies a simple transformation of the dielectric tensor, as explained in appendix F of Ref.~\onlinecite{PhysRevB.92.064403}.

\textit{High density of short wavelength magnons:}
This can be conceptualized as a random tilting of atomic magnetic moments.
Supercells containing 16-atoms were constructed for bcc Fe and fcc Ni, where both the direction and the magnitude of each magnetic moment were fixed.
The direction of each atomic moment was constrained to an axis randomly tilted from the $\hat{z}$-axis by an angle $\Theta$, hence the $z$-projection of all moments were the same and equal to $M_s \cos(\Theta)$, where $M_s$ is the corresponding saturation magnetization.
The azimuthal angles ($\phi$) of the axes were chosen in such a way that the total in-plane component of the magnetisation cancels out.
This way, the supercells are characterized by only one non-zero projection of the total moment and the symmetry of the dielectric tensor is the same as in FM state.


\subsection{Magnetic asymmetry for decreased total magnetization}\label{sec:asymdecrm}


In Fig.~\ref{fig:theo_asym_spectra}(a), (b), and (c), we present the calculated asymmetries of Fe for Stoner-like excitations, long wavelength and short wavelength magnons, respectively. Calculations for both the long wavelength and the short wavelength magnons excitations show a monotonic decrease in asymmetry signal with decreasing magnetization for all energies.
The situation is different for the case of Stoner-type excitation, which exhibits shifts and significant changes so that a monotonic decrease is no longer observed. This is particularly clear at the energies highlighted with a dashed ellipse.  
Even though the modeling including the short wavelength magnons exhibits a monotonic behavior, there are energy regions, primarily above the peak asymmetry, that show irregular variations with magnetization. The most conspicuous behaviour is highlighted with a dashed ellipse and it is clear that for these energies, the asymmetry and magnetic moment are not linearly proportional to each other. 

The overall behaviour of Fig.~\ref{fig:theo_asym_spectra}(a), (b), and (c), can be compared to the experimental asymmetry shown in Fig.~\ref{fig:Demagcurvexp}(a). 
For the remagnetization process, we find that almost all energies show an almost linear dependence between asymmetry and magnetic moments. Judging from Fig.~\ref{fig:theo_asym_spectra}, the remagnetization is consistent with magnon excitations (both short and long wavelengths) and that Stoner excitations are less frequent. 
Interestingly, the Ni asymmetries for all three types of excitation shown in Figs.~\ref{fig:theo_asym_spectra} (d), (e) and (f), exhibit no large irregular variations. As for Fe, there are non-linear contributions at high photon energies, but these asymmetries are too small to be experimentally measured. This correlates well with the measured energy independent behavior of Ni in Fig.~\ref{fig:MDT_ratio_fig}(b). Intuitively, an energy independent behavior would suggest long wavelength magnons to be the dominant excitation process, however, the experimental data is now determined to also support a decreased exchange splitting of Ni which has been found in other studies \cite{PhysRevLett.90.247201,PhysRevLett.111.167204}.

\begin{figure}
           \includegraphics[width=0.8\columnwidth]{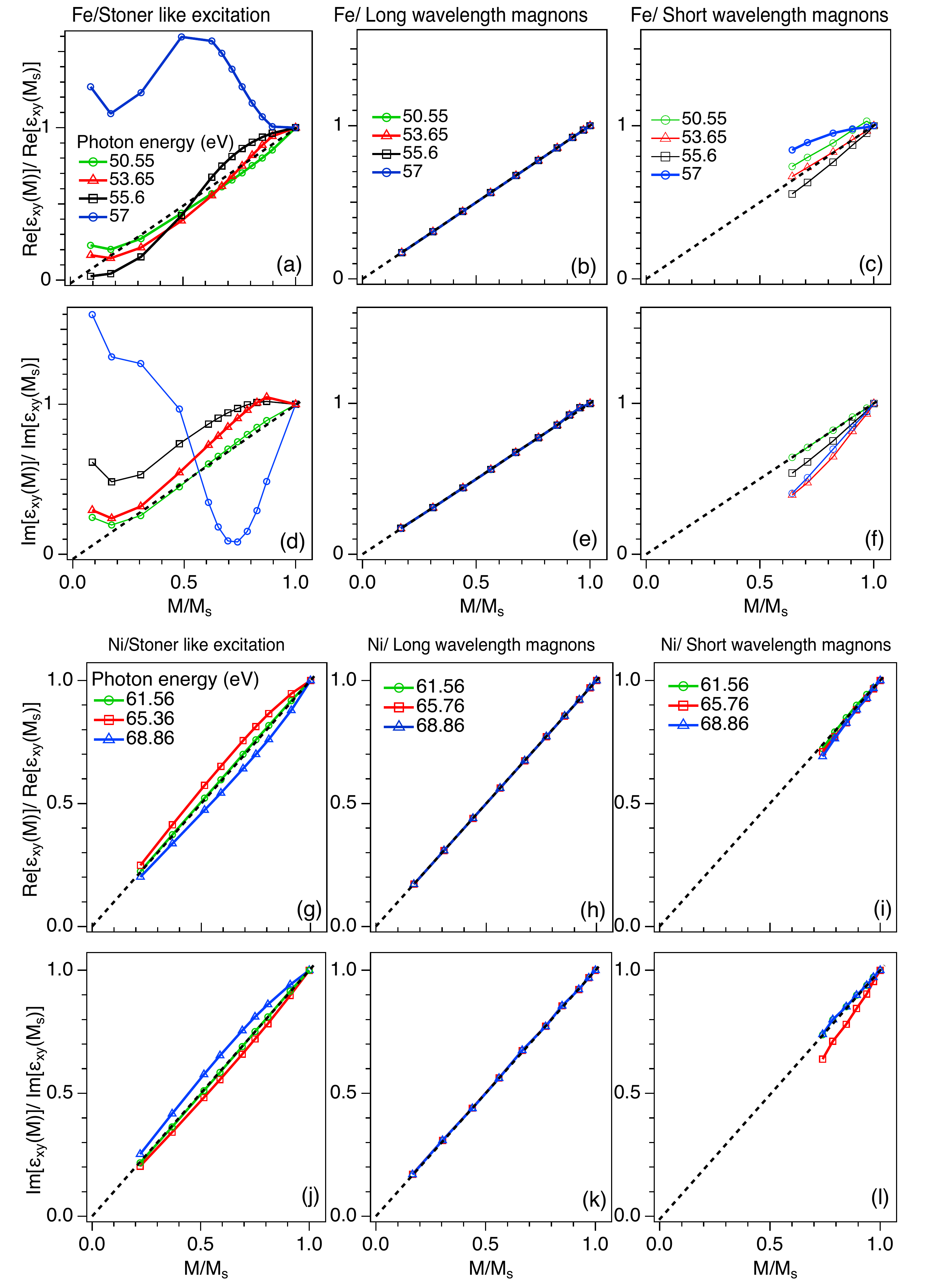}
\caption{Calculated real and imaginary parts of $\epsilon_{xy}$ as a function of reduced magnetization M/M$_s$, for the three different types of excitations at chosen energies, indicated by arrows in the top panels in Fig.~\ref{fig:theo_asym_spectra}. Results for Fe are in panels (a-f) and for Ni in (g-l). The components of $\epsilon_{xy}$ are normalized with respect to their values obtained for the ferromagnetic ground state (saturated magnetization).}
\label{fig:theo_resonance_eps}
\end{figure}


As discussed above, a proportionality between asymmetry and magnetization cannot be assumed generally for all possible types of excitations and materials. To highlight the effect of non-linearities imposed by the relation between $\epsilon_{xy}$ and magnetization, we show the  real and imaginary parts of $\epsilon_{xy}$ as a function of  magnetization at several photon energies in Fig.~\ref{fig:theo_resonance_eps} (a)-(c) and  (d)-(e) for Fe and Ni, respectively. 
Here $\epsilon_{xy}$ has been obtained at the energies indicated by the arrows in Fig.~\ref{fig:theo_asym_spectra} (a) and Fig.~\ref{fig:theo_asym_spectra}(d). 
For Ni, it is clear that an almost proportional relation exists between $\epsilon_{xy}$ and the magnetization for all three types of magnetic excitations. For Fe, there is an approximate proportionality for excitations corresponding to long and short wavelength magnons. For Stoner type excitations with a collinear decrease of the magnetic moments, there are strong deviations from the proportionality, as shown in Fig.~\ref{fig:theo_asym_spectra}. As a matter of fact, such type of non-linearities between {\bf M} and  $\epsilon_{xy}$ were predicted theoretically long time ago \cite{erskine_calculation_1975}.

\begin{figure}
           \includegraphics[width=0.8\columnwidth]{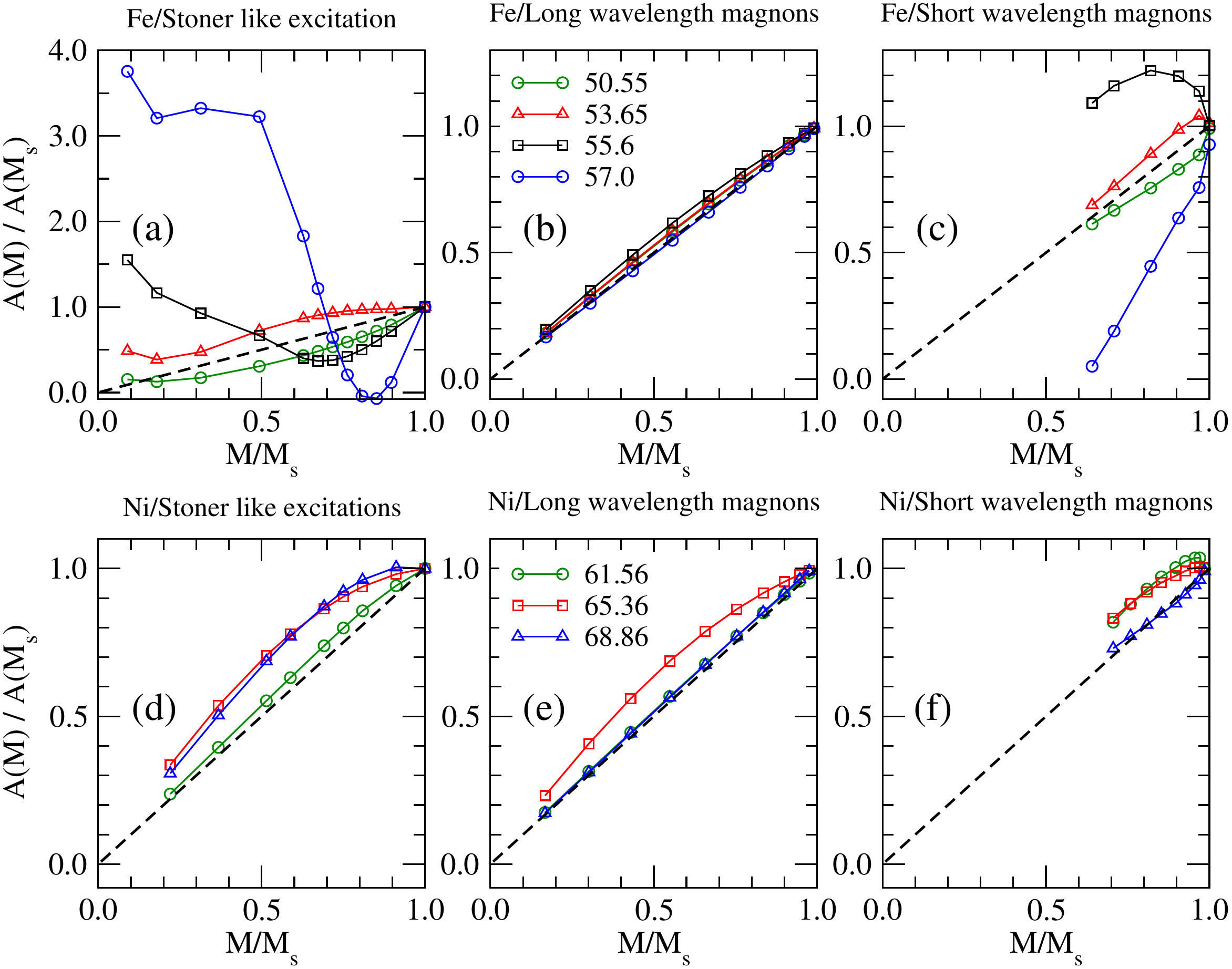}
\caption{Calculated magnetic asymmetry as a function of reduced magnetization M/M$_s$, for the three different types of excitations at chosen energies. Results for Fe are in panels (a-c) and for Ni in (d-f). The asymmetries are normalized with respect to their values obtained for the ferromagnetic ground state (saturated magnetization).}
\label{fig:theo-Anorm}
\end{figure}

Further insight can be obtained by analysing the relation between the magnetic asymmetry and magnetization in the same way.
The corresponding results are shown in Fig.~\ref{fig:theo-Anorm}
In case of Fe, one can see that the overall behaviour of asymmetry is reminiscent to that of $\epsilon_{xy}$ (Fig.~\ref{fig:theo_resonance_eps}(a-f)).
Strongly non-linear behaviour of Fe for the Stoner and short wavelength magnons, already observed for $\epsilon_{xy}$ is further enhanced.
At the same time, even though the data for Ni shows a better proportionality to the magnetization, the non-linear effects also become apparent.
Here, however, these non-linearities do not originate from the changes in the dielectric tensor, shown above in Fig.~\ref{fig:theo_resonance_eps}(g-l).
This is best seen for the long wavelength magnons, which are characterized by perfectly linear $\epsilon_{xy}$, yet exhibiting a deviation from proportionality for the magnetic asymmetry at the energies close to the main peak (65.36 eV).
Same kind of non-linearities dominate the spectra for other two types of considered magnetic excitations.
Combining the information obtained from Figs.~\ref{fig:theo_resonance_eps} and \ref{fig:theo-Anorm}, we are thus able to distinguish the non-linear effects originating from the intrinsic changes in the band structure and from the breakdown of a linear relation between asymmetry and $\epsilon_{xy}$ (Eq.~\ref{prop-rel2}).

In previous sections, we have shown that how various kinds of magnetic excitations will alter the dielectric tensor and the magnetic asymmetry. 
However, it is not immediately clear which excitations are relevant for different materials, and how they would contribute to the optical response. 
Given that the system is strongly excited, it is not expected to be in a single configuration (microstate). Rather, we expect it to be in a (partially equilibrated) thermal average over a range of excitations. 
Of all the excited states, only the ones with a sufficient population contribute to the magnetic asymmetry. These are states with an excitation energy above the ground state, that are comparable to the average energy per unit cell or the thermal energy (notice that even after a few picoseconds the system of electrons and magnons still retains a rather high effective temperature).
Thus the relative population in principle can be deduced from an analysis of the DFT total energies of the microstates, studied in the previous section.
However, a problem appears in practice due to that the non-collinear configuration which we considered for the case of simulating a high density of short wavelength magnons is limited, and does only represent a small subset of allowed configurations. Thus, we can not quantitatively compare the relative energies of these states, and compare them to e.g. the Stoner excitations. 

It is nevertheless worth analyzing results which are obtained for a restricted set of non-collinear states, keeping in mind that they do not provide the complete picture. 
Our results indicate that in order to reach a small decrease of the magnetization in bcc Fe, Stoner excitations and high density magnon excitations have similar energies. In order to reach very small values of the magnetization ($<0.8M_s$), the non-collinear configuration will be energetically preferable compared to reduced values of the moments, that would be dictated by Stoner excitations. On the other hand, Ni being an itinerant ferromagnet with a rather low value of the saturation moment, shows a completely different behaviour. Here the Stoner excitations provide the much more efficient channel for the reduction of the magnetization compared to the selected short wavelength magnons. These results are in qualitative agreement with recent calculations of the temperature dependent magnetic properties, where a different behaviour of the two systems has been reported \cite{PhysRevB.95.184432}.

\section{Discussion and Conclusion}\label{sec:discussion}

The relationship between sample magnetization and measured asymmetry, obtained from T-MOKE experiments around the M-absorption edges of Fe an Ni, has been investigated using a pump-probe technique. The measured asymmetries have been compared with those obtained from theoretical modeling, showing a good qualitative agreement. Calculations have also been used to determine the relative contributions of three types of magnetic excitations to the variations of asymmetry during the phases of demagnetization and remagnetization. 

In Ref.~\cite{Turgut2016StonerDemagnetization}, the authors suggest that Stoner and spin wave excitations give distinct fingerprints in the magneto-optical spectra, in agreement with an earlier theoretical proposal \cite{erskine_calculation_1975}.
Our results indicate that considering highly non-collinear spin states induces yet other types of non-linearities both in dielectric tensor and in magnetic asymmetry and can preclude a clear identification of the main excitation mechanism.

Experimentally, we found that Fe shows a strong energy dependence of the asymmetry during demagnetization. The origin of this behavior cannot be uniquely determined by comparing the different theoretical models to experimental data. However, it is clear that the out-of-equilibrium situation induced by the pump excitation creates an electronic and magnetic state that has a more complex magneto-optical response than the assumptions that lead to Eq.~3.
The time-independent MAT ratios, found for times $t \geq 2$ ps for Fe, indicate that the remagnetization process must be dominated by magnons in this time range. 
In contrast, for Ni the energy dependence of the asymmetry does not show qualitative differences during demagnetization and remagnetization. 
Theoretical calculations reveal that all considered magnetic excitations result in changes of the off-diagonal component of dielectric tensor, scaling linearly with the magnetization.
At the same time, we report that in Ni there is a non-linear behaviour of asymmetry with respect to $\epsilon_{xy}$, which is especially pronounced at the main peak. This type of non-linearities are not related with the modifications of the electronic structure.

Overall, the results presented here point out that a direct linear relationship between asymmetry measured in pump-probe experiments, and magnetization cannot be assumed in general. This implies that one needs to interpret these types of experiments with some care, and that ideally one should investigate the optical response for more than one photon energy of the probe. Furthermore, our results stress that a deeper insight into the nature of the magnetic excitations of a specific material can be obtained by integrating these types of experiments with a more involved theoretical analysis than usually done.

\section{acknowledgement}
The financial support from the Swedish Research Council (VR, contracts 2016-04524 and 2013-08316) is gratefully acknowledged. O.E. acknowledges the financial support from the Knut and Alice Wallenberg Foundation (KAW), The Foundation for Strategic Research, Energimyndigheten, STandUPP and eSSENCE. M.B. acknowledges Nanyang Technological University, NAP-SUG for funding. I.D.M. acknowledges support by the appointment to the JRG program at the APCTP through the Science and Technology Promotion Fund and Lottery Fund of the Korean Government, as well as support by the Korean Local Governments, Gyeongsangbuk-do Province and Pohang City.
The computational work was performed on resources provided by the Swedish National Infrastructure for Computing (SNIC) at the High Performance Computing Center North (HPC2N), at the National Supercomputer Center (NSC) and at the PDC center for High Performance Computing.

\bibliography{myrefs}

\end{document}